\documentclass[12pt]{revtex4}

\usepackage{graphicx,color}
\usepackage{epstopdf}
\usepackage{amssymb}
\begin{document}

\title{Path Integral Methods for Stochastic Differential Equations}
\author{Carson C. Chow and Michael A. Buice}
\address{Laboratory of Biological Modeling, NIDDK, NIH,
Bethesda, MD 20892}

\date{\today}

\begin{abstract}
We give a pedagogical review of the application of field theoretic and path integral methods to calculate moments of the probability density function of stochastic differential equations perturbatively.  
\end{abstract}

\maketitle

\section{Introduction}
There are many applications of stochastic differential equations (SDE) for mathematical modeling.  In the realm of neuroscience, SDEs are utilized to model stochastic phenomena that range in scale from molecular transport in neurons, to neuronal firing, to networks of coupled neurons, to even cognitive phenomena such as decision problems \cite{tuckwell:1989fk}.  In many applications, what is often desired is the ability to obtain closed form solutions or approximations of quantities such as the moments of stochastic processes.  However,  generally these SDEs are nonlinear and difficult to solve.   
For example one often encounters equations of the form
\begin{eqnarray*}
\frac{dx}{dt} =f(x) + g(x)\eta(t)
\end{eqnarray*}
where $\eta(t)$ represents some noise process, in the simplest case a white noise process where $\langle \eta(t) \rangle = 0$ and $\langle \eta(t) \eta(t') \rangle =  \delta(t - t')$.  Often of interest are the moments of $x(t)$ or the probability density function $p(x,t)$.  Traditional methods use Langevin or Fokker-Planck approaches to compute these quantities, which can still be difficult and unwieldy to apply perturbation theory \cite{risken,Gardiner:2004eu,Kampen:2007hl}.  Here, we will show how methods developed in nonequilibrium statistical mechanics using path or functional integrals  \cite{doi1,doi2,peliti,Janssen2005147,cardyrev,cardyrev2,2004piqm.book.....K,tauberbook} can be applied to solve SDEs.  While these methods have been recently applied at the level of networks, the methods are applicable to more general stochastic processes~ \cite{bc, buice-cowan,bcc,bressloffpath,Hildebrand:2006p4,buice:031118} .
%
Path integral methods provide a convenient tool to compute quantities such as moments and transition probabilities perturbatively.  They also make renormalization group methods available when perturbation theory breaks down.

Although Wiener introduced path integrals to study stochastic processes, these methods are not commonly used nor familiar to much of the neuroscience or applied mathematics community.  There are many textbooks on path integrals but most are geared towards quantum field theory or statistical mechanics~\cite{zinnjustin,Kardar:2007fk,Chaichian:2001}.  Here we give a  pedagogical review of these methods specifically applied to SDEs.  In particular, we review the response function method \cite{martin, tauber}, which is particularly convenient to compute desired quantities such as moments.  

The goal of this review is to present methods to compute actual quantities.  Thus, mathematical rigor will be dispensed for convenience.   This review will be  elementary.  In Section~\ref{sec:momgenfunc}, we cover moment generating functionals, which expand the definition of generating functions to cover distributions of functions, such as the trajectory $x(t)$ of a stochastic process. We continue in Section~\ref{sec:appsde} by constructing functional integrals appropriate for the study of SDEs, using the Ornstein-Uhlenbeck process as an example.  Section~\ref{sec:pertfeyn} introduces the concept of Feynman diagrams as a tool for carrying our perturbative expansions and introduces the ``loop expansion", a tool for constructing semiclassical approximations.  The following section \ref{sec:fp} provides the connection between SDEs and equations for the density $p(x,t)$ such as Fokker-Planck equations.  Finally, we end the paper by pointing the reader towards important entries to the literature.

\section{Moment generating functionals}
\label{sec:momgenfunc}

The strategy of path integral methods is to derive a generating function or functional for the moments and response functions for SDEs.  The generating functional will be an infinite dimensional generalization for the familiar generating function for a single random variable.  In this section we review moment generating functions and show how they can be generalized to functional distributions.

Consider a probability density function (PDF) $P(x)$ for a single real variable $x$.  The moments of the PDF are given by
\begin{eqnarray*}
\langle x^n \rangle = \int x^n P(x) dx
\end{eqnarray*}
and can be obtained directly by taking derivatives of the generating function
\begin{eqnarray*}
Z(\lambda)= \langle e^{\lambda x} \rangle = \int e^{\lambda x} P(x) dx
\end{eqnarray*}
with
\begin{eqnarray*}
\langle x^n \rangle = \frac{1}{Z[0]}\left.\frac{d^n}{d \lambda^n}Z(\lambda)\right|_{\lambda=0}
\end{eqnarray*}
Note that in explicitly including $Z[0]$ we are allowing for the possibility that $P(x)$ is not normalized.    This freedom will be convenient especially when we apply perturbation theory.

For example, the generating function for a Gaussian PDF, $P(x)\propto e^{-\frac{(x-a)^2}{2\sigma^2}}$, is
\begin{eqnarray}
Z(\lambda)=\int_{-\infty}^\infty e^{-\frac{(x-a)^2}{2\sigma^2}+\lambda x}dx
\label{gaussgen}
\end{eqnarray}
The integral can be computed by completing the square so that the exponent of the integrand can be written as a perfect square
\begin{eqnarray*}
-\frac{(x-a)^2}{2\sigma^2} +\lambda x = -A(x-x_c)^2+B
\end{eqnarray*}
This is equivalent to shifting $x$ by $x_c$, which is the critical or stationary point of the exponent
\begin{eqnarray*}
\frac{d}{dx}\left(-\frac{(x-a)^2}{2\sigma^2} +\lambda x \right)=0
\end{eqnarray*}
yielding
$ x_c=\lambda\sigma^2+a$.
The constants are then
\begin{eqnarray*}
A=\frac{1}{2\sigma^2}
\end{eqnarray*}
and
\begin{eqnarray*}
B=\frac{x_c^2}{2\sigma^2}-\frac{a^2}{2\sigma^2}=\frac{\lambda^2\sigma^2}{2}+\lambda a
\end{eqnarray*}
The integral in (\ref{gaussgen}) can then be computed to obtain
\begin{eqnarray*}
Z(\lambda)=\int_{-\infty}^\infty e^{-\frac{(x-\lambda\sigma^2-a)^2}{2\sigma^2}+\lambda a+\frac{\lambda^2\sigma^2}{2}}dx=Z(0)e^{\lambda a+\frac{\lambda^2\sigma^2}{2}}
\end{eqnarray*}
where
\begin{eqnarray*}
Z(0)=\int_{-\infty}^\infty e^{-\frac{x^2}{2\sigma^2}}dx = \sqrt{2\pi}\sigma
\end{eqnarray*}
is a normalization factor.  The mean of $x$ is then given by
\begin{eqnarray*}
\langle x \rangle=\frac{d}{d\lambda} \left.e^{\lambda a+\frac{\lambda^2\sigma^2}{2}}\right |_{\lambda=0}=a
\end{eqnarray*}


The cumulant generating function is defined as
\begin{eqnarray*}
W(\lambda)=\ln Z(\lambda)
\end{eqnarray*}
so that the cumulants are
\begin{eqnarray*}
\langle x^n \rangle_C=\left.\frac{d^n}{d\lambda^n} W(\lambda)\right|_{\lambda=0}
\end{eqnarray*}
In the Gaussian case
\begin{eqnarray*}
W(\lambda)=\lambda a + \frac{1}{2}\lambda^2\sigma^2+\ln Z(0)
\end{eqnarray*}
yielding
$\langle x \rangle_C =\langle x \rangle= a$ and $\langle x^2 \rangle_C\equiv {\rm var}(x)=\langle x^2 \rangle -\langle x\rangle^2= \sigma^2$,
and $\langle x^n\rangle_C=0$, $n>2$.

The generating function can be generalized for an n-dimensional vector $x=\{x_1,x_2,\cdots,x_n\}$ to become a  generating functional that maps the $n$-dimensional vector $\lambda=\{\lambda_1,\lambda_2,\dots,\lambda_n\}$ to a real number with the form
\begin{eqnarray*}
Z[\lambda]=\int\prod_{i=1}^n dx_i e^{-\frac{1}{2}\sum_{j,k} x_jK^{-1}_{jk}x_k+\sum_j \lambda_j x_j}
\end{eqnarray*}
where $ K^{-1}_{jk}\equiv  (K^{-1})_{jk}$ and we use square brackets to denote a functional.
This integral can be solved by transforming to orthonormal coordinates, which is always possible if $K^{-1}_{ij}$ is symmetric, as it can be assumed to be.  Hence, let $\omega_\alpha$ and $v^\alpha$ be the $\alpha$th eigenvalues and orthonormal eigenvectors of $K^{-1}$ respectively, i.e.
\begin{eqnarray*}
\sum_j K^{-1}_{ij} v_j^\alpha = \omega_\alpha v_i^\alpha
\end{eqnarray*}
and
\begin{eqnarray*}
\sum_j v_j^\alpha v_j^\beta=\delta_{\alpha\beta}
\end{eqnarray*}
Now, expand $x$ and $\lambda$ in terms of the eigenvectors with
\begin{eqnarray*}
x_k=\sum_\alpha  c_\alpha v^{\alpha}_k
\end{eqnarray*}
\begin{eqnarray*}
\lambda_k=\sum_\alpha  d_\alpha v^{\alpha}_k
\end{eqnarray*}
Hence
\begin{eqnarray*}
\sum_{j,k} x_jK^{-1}_{jk} x_k=\sum_{j}\sum_{\alpha,\beta} c_\alpha \omega_\beta c_\beta v_j^\alpha v_j^\beta=\sum_{\alpha,\beta} c_\alpha \omega_\beta c_\beta \delta_{\alpha\beta}=\sum_\alpha \omega_\alpha c^2_\alpha
\end{eqnarray*}
Since, the  Jacobian is $1$ for an orthonormal transformation the generating functional is
\begin{eqnarray*}
Z[\lambda]&= &\int \prod_\alpha dc_\alpha e^{\sum_{\alpha}(-\frac{1}{2}\omega_\alpha c^2_\alpha + d_\alpha c_\alpha)}\\
&=&\prod_\alpha \int_{-\infty}^\infty dc_\alpha e^{-\frac{1}{2}\omega_\alpha c^2_\alpha + d_\alpha c_\alpha}\\
&=&Z[0]\prod_\alpha  e^{\frac{1}{2}\omega_\alpha^{-1} d^2_\alpha }\\
&=&Z[0]e^{\sum_{jk}\frac{1}{2} \lambda_j K_{jk}\lambda_k }
\end{eqnarray*}
where
\begin{eqnarray*}
Z[0]=(2\pi\det{K})^{n/2}
\end{eqnarray*}
The cumulant generating functional is
\begin{eqnarray*}
W[\lambda]=\ln Z[\lambda]
\end{eqnarray*}
Moments are given by



\begin{eqnarray*}
\left \langle \prod_{i=1}^{s} x_i \right \rangle = \left.\frac{1}{Z[0]}\prod_{i=1}^{s}\frac{\partial}{\partial \lambda_i} Z[\lambda]\right|_{\lambda_i=0}
\end{eqnarray*}
However,  since the exponent is quadratic in the components $\lambda_l$, only even powered moments are non-zero.  From this we can deduce that
\begin{eqnarray*}
\left \langle \prod_{i=1}^{2s} x_i \right \rangle=\sum_{\rm all \ possible \ pairings} K_{i_1,i_2}\cdots K_{i_{2s-1}i_{2s}}
\end{eqnarray*}
which is known as Wick's theorem.  Any Gaussian moment about the mean can be obtained by taking the sum of all the possible ways of ``contracting" two of the variables.
For example
\begin{eqnarray*}
\langle x_a x_b x_c x_d\rangle= K_{ab}K_{cd} +K_{ad}K_{bc}+K_{ac}K_{bd}
\end{eqnarray*}
%

In the continuum limit,  a generating functional for a function $x(t)$ on the real domain $t\in [0,T]$ is obtained by taking a limit of the generating functional for the vector  $x_i$.  Let the interval $[0,T]$ be divided into $n$ segments of length $h$ so that $T=nh$ and 
$x(t/h)=x_i$ for $t\in[0,T]$.   We then take the limit of $n\rightarrow \infty$ and $h\rightarrow 0$ preserving $T= n h$.
We similarly identify $\lambda_i\rightarrow \lambda(t)$ and $K_{ij}\rightarrow K(s,t)$ 
and obtain
\begin{eqnarray*}
Z[\lambda]=\int{\cal D}x(t) e^{-\frac{1}{2}\int x(s) K^{-1}(s,t) x(t) ds dt +\int \lambda(t)x(t) dt}
\end{eqnarray*}
\begin{eqnarray}
= Z[0]e^{\int\frac{1}{2} \lambda(s)K(s,t)\lambda(t) ds dt }
\label{eq:freegenfunc}
\end{eqnarray}
where the  the measure for integration
\begin{eqnarray*}
{\cal D} x(t)\equiv \lim_{n\rightarrow\infty}\prod_{i=0}^n dx_i 
\end{eqnarray*}
is over functions.
Although $Z[0]=\lim_{n\rightarrow\infty}(2\pi\det K)^{n/2}$ is formally infinite, the moments of the distributional are well defined.  The integral is called a path integral  or  a functional integral.  Note that $Z[\lambda]$ refers to a functional that maps different ``forms" of the function $\lambda(t)$ over the time domain to a real number.
Defining the functional derivative to obey all the rules of the ordinary derivative with
\begin{eqnarray*}
\frac{\delta \lambda(s)}{\delta\lambda(t)}=\delta(s-t)
\end{eqnarray*}
the moments again obey
\begin{eqnarray*}
\left \langle \prod_i x(t_i) \right \rangle = \frac{1}{Z[0]}\prod_i \frac{\delta}{\delta\lambda(t_i)}Z[\lambda]
\end{eqnarray*}
\begin{eqnarray*}
=\sum_{\rm all \ possible \ pairings} K(t_{i_1},t_{i_2})\cdots K(t_{i_{2s-1}},t_{t_{i_{2s}}})
\end{eqnarray*}
For example
\begin{eqnarray*}
\langle x(t_1)x(t_2)\rangle = \frac{1}{Z[0]}\frac{\delta}{\delta\lambda(t_1)}
\frac{\delta}{\delta\lambda(t_2)}Z[\lambda]=K(t_1,t_2)
\end{eqnarray*}
We can further generalize the generating functional to describe the probability distribution of a function $\varphi(\vec{x})$ of a real vector $\vec{x}$, instead of a single variable $t$ with
\begin{eqnarray*}
Z[\lambda]&=&\int{\cal D}\varphi e^{-\frac{1}{2}\int \varphi(\vec{y}) K^{-1}(\vec{y},\vec{x}) \varphi(\vec{x}) d^dy d^dx +\int \lambda(\vec{x})\varphi(\vec{x}) d^dx}\\
&=& Z[0]e^{\int\frac{1}{2} \lambda(\vec{y})K(\vec{y},\vec{x})\lambda(\vec{x}) d^dy d^dx }
\end{eqnarray*}
Historically, computing moments and averages of a probability density functional of a function of more than one variable is called field theory.
In general, the probability density functional is usually written in exponential form
\begin{eqnarray*}
P[\varphi]=e^{-S[\varphi(\vec{t})]}
\end{eqnarray*}
where $S[\varphi]$ is called the action and the generating functional is often written as
\begin{eqnarray*}
Z[J]=\int {\cal D}\varphi e^{-S[\phi] + J\cdot \varphi}
\end{eqnarray*}
where
\begin{eqnarray*}
J\cdot \varphi = \int J(\vec{t})\varphi(\vec{t}) d^dt
\end{eqnarray*}
For example, the action given by
\begin{eqnarray*}
S[\varphi] = \int \varphi(\vec{t}) K^{-1}(\vec{t},\vec{t'})\varphi(\vec{t'}) d^dt d^dt' + g\int \varphi^4(\vec{t}) d^dt\end{eqnarray*}
is called $\varphi^4$ (``$\varphi$-4") theory.

The analogy between stochastic systems and quantum theory, where path integrals are commonly used, is seen by transforming the time coordinates in the path integrals  via $t \rightarrow it$ (where $i^2 = -1$).  When the field $\phi$ is a function of a single variable $t$, then this would be analogous to single particle quantum mechanics where the quantum amplitude can be expressed in terms of a path integral over a configuration variable $\phi(t)$.  When the field is a function of two or more variables $\phi(\vec{r},t)$, then this is analogous to quantum field theory, where the quantum amplitude is expressed as a path integral over the quantum field $\phi(\vec{r},t)$.

\section{Application to SDE}
\label{sec:appsde}

Building on the previous section, here we derive a generating functional for  SDEs.
Consider a Langevin equation
\begin{eqnarray*}
\frac{dx}{dt}=f(x,t)+ g(x,t)\eta(t)
\label{sde}
\end{eqnarray*}
with initial condition $x(t_0)=y$, on the domain $t\in [0,T]$.  
 Equation (\ref{sde})
is to be interpreted as the Ito stochastic differential equation
\begin{eqnarray}
dx=f(x,t)dt+ g(x,t)dB_t
\label{ito}
\end{eqnarray}
where $dB_t$ is a Brownian stochastic process.   We will show how to generalize to other stochastic processes later.
According to the convention for an Ito stochastic process, $g(x,t)$ is  \emph{non-anticipating}, which means that in evaluating the integrals over time and $B_t$, $g(x,t)$ is independent of $B_\tau $ for $\tau > t$.   The choice between Ito and Stratonovich conventions amounts to a choice of the measure for the path integrals, which will be manifested in a condition on the linear response or ``propagator" that we introduce below.  

The goal is to derive a probability density functional (PDF) and moment generating functional for the stochastic variable $x(t)$.
For the path integral formulation, it is more convenient to take $x(t_0)=0$ in (\ref{sde}) and enforce the initial condition with a source term so that
\begin{eqnarray}
\frac{dx}{dt}=f(x,t)+ g(x,t)\eta(t) + y\delta(t-t_0)
\label{sde2}
\end{eqnarray}
where $\delta(\cdot)$ is the point mass or Dirac delta functional. 
The discretized form of (\ref{sde2}) with the Ito interpretation for small time step $h$ is given by
\begin{eqnarray}
x_{i+1}-x_i = f_i(x_i)h + g_i(x_i)w_i\sqrt{h} +y\delta_{i,o}
\label{discrete}
\end{eqnarray}
$i\in \{0,1,\dots,N\}$, $T= Nh$, $\delta_{i,j}$ is the Kronecker delta, $x_0=0$, and
 $w_i$ is a discrete random variable with $\langle w_i\rangle=0$ and $\langle w_iw_j\rangle=\delta_{i,j}$. Hence, the discretized stochastic variable vector $x_i$ depends on the discretized white noise process $w_i$ and the initial condition $x_0$.  We use $x$ and $w$ without indices to denote the vectors $x = (x_1, \dots, x_N)$ and $w = (w_0, w_1, \dots, w_{N-1})$.  
   Formally, the joint PDF for the vector $x$ can be written as
\begin{eqnarray*}
P[x | w; y]= \prod_{i=0}^N \delta[x_{i+1}-x_i - f_i(x_i)h - g_i(x_i)w_i\sqrt{h} -y\delta_{i,0}]
\end{eqnarray*}
i.e. the probability density function is given by the point mass (Dirac delta) constrained at the solution of the SDE.

Inserting the Fourier representation of the Dirac delta 
\begin{eqnarray*}
\delta(z_i) =\frac{1}{2\pi} \int e^{-ik_i z_i} dk_i
\end{eqnarray*}
gives
\begin{eqnarray*}
P[x | w;y]=\int\prod_{j=0}^N \frac{dk_j}{2\pi}e^{-i\sum_j k_j(x_{j+1}-x_j-f_j(x_j)h-g_j(x_j)w_j\sqrt{h}-y\delta_{j,0})}
\end{eqnarray*}
The PDF is now expressed in exponential form.

For Gaussian white noise the PDF of $w_i$ is given by
\begin{eqnarray*}
P(w_i)=\frac{1}{\sqrt{2\pi}} e^{-\frac{1}{2}w_i^2}
\end{eqnarray*}
Hence

\begin{eqnarray*}
P[x | y]&=&\int P[x | w; y] \prod_{j=0}^N P(w_j) dw_j  \\
&=&\int\prod_{j=0}^N \frac{dk_j}{2\pi} e^{-i\sum_j k_j(x_{j+1}-x_j-f_j(x_j)h - y\delta_{j,0})}\int \prod_{j=0}^N \frac{dw_j}{\sqrt{2\pi}} e^{ik_jg_j(x_j)w_j\sqrt{h}}e^{-\frac{1}{2}w_j^2}
\end{eqnarray*}
can be integrated by completing the square as demonstrated in the previous section to obtain
\begin{eqnarray*}
P[x | y]=\int\prod_{j=0}^N \frac{dk_j}{2\pi} e^{-\sum_j (ik_j) \left (\frac{x_{j+1}-x_j}{h}-f_j(x_j)-y\frac{\delta_{j,0}}{h} \right )h+\sum_j \frac{1}{2} g_j^2(x_j)(ik_j)^2h}\end{eqnarray*}
Taking the continuum limit $h\rightarrow 0$, $N\rightarrow\infty$ such that $T=Nh$ gives
\begin{eqnarray*}
P[x(t) | y,t_0]=\int{\cal D}\tilde{x}(t) e^{-\int  \left [  \tilde{x}(t)(\dot{x}(t)-f(x(t),t)-y\delta(t-t_0)) - \frac{1}{2}\tilde{x}^2(t) g^2(x(t),t) \right ] dt}
\end{eqnarray*}
with a newly defined complex variable $i k_i \rightarrow \tilde{x}(t)$. We include the argument of $x(t)$ as a reminder that this is a functional of $x$ conditioned on two scalars $y$ and $t_0$.  The moment generating functional for $x(t)$ and $\tilde{x}(t)$ is then given by
\begin{eqnarray*}
Z[J,\tilde{J}]=\int {\cal D} x(t){\cal D}\tilde{x}(t) e^{-S[x,\tilde{x}] +\int \tilde{J}(t)x(t) dt + \int J(t) \tilde{x}(t) dt}
\end{eqnarray*}
with action
\begin{eqnarray}
S[x,\tilde{x}]= \int  \left [ \tilde{x}(t)(\dot{x}(t)-f(x(t),t) -y\delta(t-t_0)) - \frac{1}{2}\tilde{x}^2(t) g^2(x(t),t) \right ] dt
\label{actionSDE}
\end{eqnarray}

The probability density functional can be derived directly from the SDE (\ref{sde}) by considering the infinite dimensional Dirac delta functional and taking the path integral:
\begin{eqnarray*}
P[x(t) | y, t_0]&=&\int {\cal D}\eta(t) \delta[\dot{x}(t)-f(x,t)-g(x,t)\eta(t)-y\delta(t-t_0)]e^{-\int \eta^2(t) dt}\\
&=&\int {\cal D}\eta(t){\cal D}\tilde{x}(t) e^{-\int \tilde{x}(t) (\dot{x}(t)-f(x,t)-y\delta(t-t_0)) +\tilde{x}(t)g(x,t)\eta(t)- \eta^2(t) dt}\\
&=&\int {\cal D}\tilde{x}(t) e^{-\int \tilde{x}(t)(\dot{x}(t)-f(x,t)-y\delta(t-t_0)) +\frac{1}{2}\tilde{x}^2(t)g^2(x,t) dt}
\end{eqnarray*}
yielding the action (\ref{actionSDE}) \footnote{This derivation is, strictly speaking, incorrect because the delta functional fixes the value of $\dot{x}(t)$, not $x(t)$.  It works because the Jacobian under a change of variables from $\dot{x}(t)$ to $x(t)$ is $1$.}. Owing to the definition $i k_i \rightarrow \tilde{x}(t)$ the integrals over $\tilde{x}(t)$ are along the \emph{imaginary} axis, which is why no explicit $i$ appears in the action above.




In a similar manner, we can define the path integral for more general processes than the Brownian motion processes that we are using.  Let $\eta(t)$ instead be a process with  cumulant generating functional $W[\lambda(t)]$ so that the cumulants of $\eta(t)$ (which may depend upon $x(t)$) are given by functional derivatives with respect to $\lambda(t)$.  This process will have its own action $S[\eta(t)]$ and the path integral can be written as
\begin{eqnarray*}
P[x(t) | y, t_0]&=&\int {\cal D}\eta(t) \delta[\dot{x}(t)-f(x,t)-\eta(t)-y\delta(t-t_0)]e^{-S[\eta(t)]}\\
&=&\int {\cal D}\eta(t){\cal D}\tilde{x}(t) e^{-\int \tilde{x}(t) (\dot{x}(t)-f(x,t)-y\delta(t-t_0)) +\tilde{x}(t)\eta(t)dt -S[\eta(t)]}
\end{eqnarray*}
Noting that
\begin{eqnarray*}
	\int {\cal D}\eta(t)e^{\int \tilde{x}(t)\eta(t)dt -S[\eta(t)]} = e^{W[\tilde{x}(t)]}
\end{eqnarray*}
is the definition of the cumulant generating functional for $\eta(t)$, we have that the path integral can be written as
\begin{eqnarray*}
P[x(t) | y, t_0]&=&\int {\cal D}\eta(t){\cal D}\tilde{x}(t) e^{-\int \tilde{x}(t) (\dot{x}(t)-f(x,t)-y\delta(t-t_0))dt +W[\tilde{x}(t)]}
\end{eqnarray*}
 In the cases where the input $\eta(t)$ is delta-correlated in time, we obtain
\begin{eqnarray*}
	W[\tilde{x}(t)] = \sum_{n=1}^{\infty}\int  g_n(x(t)) \tilde{x}(t)^n  dt=  \sum_{n=1, m=0}^{\infty}\frac{v_{nm}}{n!} \int  \tilde{x}^n(t) x^m(t) dt
\end{eqnarray*}
where we have Taylor expanded the functions $g_n(x)$.  For example, the Ito process above gives
\begin{eqnarray*}
	W[\tilde{x}(t)] = \frac{D}{2} \int \tilde{x}(t)^2 dt
\end{eqnarray*}
i.e. $v_{20} = D$ and all other $v_{nm} = 0$.

\subsection{Ornstein-Uhlenbeck Process}
Consider the Ornstein-Uhlenbeck process 
\begin{eqnarray*}
\dot{x}(t)+ax(t) -\sqrt{D}\eta(t) =0
\end{eqnarray*}
with initial condition $x(0)=y$.  
The action is
\begin{eqnarray*}
S[x,\tilde{x}]=\int \left [ \tilde{x}(t)\left (\dot{x}(t)+ax(t) - y\delta(t-t_0) \right ) -\frac{D}{2}\tilde{x}^2(t) \right ] dt
\end{eqnarray*}
Defining an inverse propagator
\begin{eqnarray*}
G^{-1}(t-t')  = \left (\frac{d}{dt} + a \right )\delta(t-t')
\end{eqnarray*}
the  action is
\begin{eqnarray*}
S[x,\tilde{x}]=\int \tilde{x}(t)G^{-1}(t-t')x(t') dtdt'-\int y\tilde{x}(t)\delta(t-t_0) dt-\int \frac{D}{2}\tilde{x}(t)^2 dt
\end{eqnarray*}
and the generating functional is
\begin{eqnarray*}
Z[J,\tilde{J}]=\int {\cal D} x(t) {\cal D}\tilde{x}(t) e^{-S[x,\tilde{x}] +\int \tilde{J}(t)x(t) dt + \int J(t) \tilde{x}(t) dt}
\end{eqnarray*}
This path integral can be evaluated directly as a Gaussian integral since the action is quadratic.  In fact integrating by $\tilde{x}$(t) gives the Onsager-Machlup path integral~\cite{risken,Chaichian:2001}, which will have a Jacobian factor depending upon whether we use Ito, Stratonovich, or some other convention for our SDE.  With the Ito convention, this Jacobian is $1$. However,  the generating functional can also be evaluated by expanding the exponent around the ``free" action given by $S_F[x(t), \tilde{x}(t)]=\int \tilde{x}(t) G^{-1}(t - t') x(t') dt dt'$.   We will demonstrate this method  since it forms the basis for perturbation theory for non-quadratic actions.  Expand the integrand of the generating functional as
\begin{eqnarray}
Z[J,\tilde{J}]=\int {\cal D} x(t){\cal D}\tilde{x}(t) e^{-\int dtdt' \tilde{x}(t)G^{-1}(t-t')x(t')} \left (1 +\mu+\frac{1}{2!}\mu^2+\frac{1}{3!}\mu^3+\cdots \right )
\label{OUGF}
\end{eqnarray}
where
\begin{eqnarray*}
\mu=y\int \tilde{x}(t)\delta(t-t_0) dt+ \int\frac{D}{2}\tilde{x}^2(t) dt +\int \tilde{J}(t)x(t) dt +\int J(t)\tilde{x}(t) dt
\end{eqnarray*}
The generating functional is now expressed as a sum of moments of the free action, which are calculated from the free generating functional
\begin{eqnarray}
Z_F[J,\tilde{J}]=\int {\cal D} x(t){\cal D}\tilde{x}(t) e^{-\int dtdt' \tilde{x}(t)G^{-1}(t-t')x(t') +\int \tilde{x}(t)J(t) dt + \int x(t) \tilde{J}(t) dt}
\label{freeGF}
\end{eqnarray}
Although this integral is similar to~(\ref{eq:freegenfunc}), there are sufficient differences to warrant an explicit computation.  We note again that $\tilde{x}$ is an imaginary variable so this integral corresponds to computing a functional complex Gaussian in two fields.

The free generating functional (\ref{freeGF}) can be integrated by discretizing and expanding in terms of orthogonal eigenfunctions as before so that 
\begin{eqnarray*}
Z_{F}[\hat{J},\hat{\tilde{J}}]=\int {\cal D} \hat{x}{\cal D}\hat{k} e^{-\sum_\omega \ i\hat{k}\hat{x}\lambda +i\hat{k}\hat{J}+ \hat{x} \hat{\tilde{J}}}
\equiv\prod_\omega \int \frac{d\hat{x}d\hat{k}}{2\pi} e^{- i\hat{k}\hat{x}\lambda +i\hat{k}{J}+ \hat{x} {\tilde{J}}}
\end{eqnarray*}
where we have set $\tilde{x} = i \hat k$ and $\int dt' G^{-1} x \rightarrow \lambda(\omega) \hat{x}(\omega)$.  The integral can be completed easily by noting that $ \int \frac{d\hat{x}d\hat{k}}{2\pi} e^{- i\hat{k}(\hat{x}\lambda -\hat{J})+ \hat{x} \hat{\tilde{J}}}= \int d\hat{x} \delta(\hat{x}\lambda-\hat{J}) e^{ \hat{x} {\tilde{J}}}=\prod_\omega  e^{\lambda^{-1} {J} {\tilde{J}}}$. Transforming back to the original coordinates gives
\begin{eqnarray}
Z_{F}[J,\tilde{J}]=e^{\int \tilde{J}(t)G(t,t'){J}(t')}
\label{freegen}
\end{eqnarray}
where $G(t,t')$ is the operator inverse of $G^{-1}(t,t')$, i.e. 
\begin{eqnarray*}
	\int dt'' G^{-1}(t,t'')G(t'',t') = \left ( \frac{d}{dt} + a \right ) G(t,t') = \delta( t - t') 
\end{eqnarray*}
Therefore
\begin{eqnarray*}
	G(t,t') = H(t - t') e^{-a(t-t')}
\end{eqnarray*}
where $H(t)$ is the left continuous Heaviside step function (i.e.  $H(0)=0$, $\lim_{t\rightarrow 0^+} H(t)=1$ and thus $\lim_{t_1\rightarrow t_2^+} G(t_1,t_2)=1$, $G(t,t)=0$).  The choice of  $H(0)=0$ is consistent with the Ito condition for the SDE and insures that the configuration variable $x(t)$ is uncorrelated with future values of the stochastic driving term.  Other choices for $H(0)$ represent other forms of stochastic calculus (e.g. $H(0) = 1/2$ is the choice consistent with Stratonovich calculus)~\footnote{This is also a manifestation of the normal-ordering convention chosen for the theory.  Zinn-Justin\cite{zinnjustin} refers to this as the ``$\epsilon(0)$ problem".}.
The free moments are given by
\begin{eqnarray*}
\left\langle \prod_{ij} x(t_i) \tilde{x}(t_j)\right\rangle_F =\prod_{ij} \left.\frac{\delta}{\delta\tilde{J}(t_i)}\frac{\delta}{\delta J(t_j)}e^{\int \tilde{J}(t)G(t,t')J(t') dtdt'}\right|_{J=\tilde{J}=0}
\end{eqnarray*}
since $Z_F[0,0]=1$.  We use a subscript $F$ to denote expectation values with respect to the free action. 
From the action of (\ref{freegen}), it is clear the nonzero free moments must have equal numbers of $x(t)$ and $\tilde{x}(t)$ due to Wick's theorem, which applies here for contractions between $x(t)$ and $\tilde x(t)$.  For example, one of the fourth moments is given by
\begin{eqnarray*}
\langle x(t_1)x(t_2) \tilde{x}(t_3)\tilde{x}(t_4)\rangle_F = G(t_1,t_3)G(t_2,t_4)+G(t_1,t_4)G(t_2,t_3)
\end{eqnarray*}

Now the generating functional for the OU process  (\ref{OUGF}) can be evaluated. The only surviving terms in the expansion will have equal numbers of $x(t)$ and $\tilde{x}(t)$.  Thus only terms with factors of 
$\int\tilde{x}(t_0)\tilde{J}(t_1) x(t_1) dt_1$, $(D/2)\int\tilde{x}^2(t_1) \tilde{J}^2(t_2)x^2(t_2)dt_1dt_2$ and $\int\tilde{J}(t_1)x(t_1) J(t_2)\tilde{x}(t_2) dt_1dt_2$ (and combinations of the three) will survive.  For the OU process, the entire series is summable.  First consider the case where $D=0$.  Because there must be equal numbers of $\tilde{x}(t)$ and $x(t)$ factors in any non-zero moment due to Wick's theorem, in this case the generating functional has the form
\begin{eqnarray}
Z
&=&1 + \sum_{m=1} \frac{1}{m! m!}\int  \left\langle \prod_{i,j=1}^m\tilde{J}(t_i)x(t_i)\tilde{x}(t_j)[y\delta(t_j-t_0) +J(t_j) ] \right\rangle_F \,\prod_{i,j=1}^mdt_i dt_j
\label{term1}
\end{eqnarray}
 From Wick's theorem, the free expectation value in (\ref{term1}) will be a sum over all possible contractions between $x(t)$ and $\tilde{x}(t)$ leading to $m!$ combinations.  Thus (\ref{term1}) is
\begin{eqnarray*}
Z=\sum_{m=1} \frac{1}{m!}\left( y\int\tilde{J}(t_1)G(t_1,t_0) dt_1+\int \tilde{J}(t') J(t'') G(t',t'') dt' dt''\right)^m
\end{eqnarray*}
which means the series is an exponential function.
The other term in the exponent of (\ref{gfOU}) can be similarly calculated  resulting in\begin{eqnarray}
Z[J(t),\tilde{J}(t)]
= \exp\left(y\int\tilde{J}(t_1)G(t_1,t_0) dt_1+\int \tilde{J}(t_1) J(t_2)  G(t_1,t_2) dt_1 dt_2\right.  \nonumber\\
\left. +\frac{D}{2}\int\tilde{J}(t_1)\tilde{J}(t_2) G(t_1,t'')G(t_2,t'') dt''dt_1 dt_2 \right)
\label{gfOU}
\end{eqnarray}
The cumulant generating functional is
\begin{eqnarray}
W[J(t),\tilde{J}(t)]=y\int\tilde{J}(t)G(t,t_0) dt+ \int \tilde{J}(t') J(t'')  G(t',t'') dt' dt'' \nonumber\\
+\frac{D}{2}\int\tilde{J}(t')\tilde{J}(t'') G(t',t)G(t'',t) dtdt' dt'' 
\label{cgfOU}
\end{eqnarray}
The only nonzero cumulants are the mean
\begin{eqnarray*}
	\langle x(t) \rangle = yG(t, t_0) 
\end{eqnarray*}
the response function 
\begin{eqnarray*}
\left\langle  x(t_1) \tilde{x}(t_2)\right\rangle_C = \frac{\delta}{\delta\tilde{J}(t_1)}\frac{\delta}{\delta J(t_2)}W[J,\tilde{J}]_{J=\tilde{J}=0} = G(t_1,t_2)  
\end{eqnarray*}
and covariance
\begin{eqnarray*}
\langle x(t_1)x(t_2)\rangle_C&\equiv&\langle x(t_1)x(t_2)\rangle- \langle x(t_1)\rangle\langle x(t_2)\rangle\\
 &=& \frac{\delta}{\delta\tilde{J}(t_1)}\frac{\delta}{\delta \tilde{J}(t_2)}W[J,\tilde{J}]_{J=\tilde{J}=0} \\
&=&D\int G(t_1,t)G(t_2,t)dt
\end{eqnarray*}

Closed form expressions for  the cumulants are obtained by using the solution for the propagator $G$.
Hence, the mean is
\begin{eqnarray}
\left\langle  x(t) \right\rangle= ye^{-a(t-t_0)}H(t-t_0)
\label{meanOU}
\end{eqnarray}
the response function is
\begin{eqnarray*}
\left\langle  x(t_1) \tilde{x}(t_2)\right\rangle= e^{-a(t_1-t_2)}H(t_1-t_2)
\end{eqnarray*}
and the covariance is
\begin{eqnarray*}
\langle x(t_1)x(t_2)\rangle_C=D\int_{t_0}^{t_2} e^{-a(t_1-t')}e^{-a(t_2-t')} H(t_1-t')H(t_2-t') dt'
\end{eqnarray*}
For $t_2\ge t_1\ge t_0$
\begin{eqnarray*}
\langle x(t_1)x(t_2)\rangle_C=D\frac{e^{2a(t_1-t_2)}-e^{-a(t_1+t_2-2t_0)}}{2a}
\end{eqnarray*}
For $t_1=t_2=t$
\begin{eqnarray}
\langle x(t)^2\rangle_C=\frac{D}{2a}(1-e^{-2at})
\label{varOU}
\end{eqnarray}

\section{Perturbative methods and Feynman diagrams}
\label{sec:pertfeyn}

If the SDE is nonlinear, then the generating functional cannot be computed exactly as in the linear case.  However, propagators and moments can be computed perturbatively.  The method we use is an infinite dimensional generalization of Laplace's method for finite dimensional integrals~\cite{bender}.  In fact, the method was used to compute the generating functional for the Ornstein-Uhlenbeck process.  The only difference is that for nonlinear SDEs the resulting asymptotic series is not generally summable.

 The strategy is again to split the action $S[x,\tilde{x}]= S_F + S_I$, where $S_F$ is called the ``free" action and $S_I$ is called the ``interacting" action.  The generating functional is
\begin{eqnarray}
Z[J,\tilde{J}]=\int {\cal D} x{\cal D}\tilde{x} e^{-S[x,\tilde{x}] +\int \tilde{J}x dt + \int J \tilde{x} dt}
\label{eq:genFuncPert}
\end{eqnarray}
The moments satisfy
\begin{eqnarray}
\left\langle \prod_i^m\prod_{j}^{n}x(t_i)\tilde{x}(t_j)\right\rangle =
\frac{1}{Z[0,0]}\prod_i^m\prod_{j}^{n} \left.\frac{\delta}{\delta\tilde{J}(t_i)}\frac{\delta}{\delta J(t_j)}  Z \right|_{J=\tilde{J}=0}
\label{moment1}
\end{eqnarray}
and the cumulants satisfy
\begin{eqnarray}
\left\langle \prod_i^m\prod_{j}^{n}x(t_i)\tilde{x}(t_j)\right\rangle_C =\prod_i^m\prod_{j}^{n} \left.\frac{\delta}{\delta\tilde{J}(t_i)}\frac{\delta}{\delta J(t_j)} \ln Z \right|_{J=\tilde{J}=0}
\label{cumulant1}
\end{eqnarray}
The generating functional is computed perturbatively by expanding the integrand of (\ref{eq:genFuncPert})  around the free action
\begin{eqnarray*}
Z[J,\tilde{J}]=\int {\cal D} x{\cal D}\tilde{x} e^{-S_F[x,\tilde{x}] }\left ( 1 + S_I+\int \tilde{J}x dt + \int J \tilde{x} dt \right. \\
\left. + 
\frac{1}{2!} \left (S_I+\int \tilde{J}x dt + \int J \tilde{x} dt \right )^2 +  \frac{1}{3!}  S_I^3+\cdots\right)
\end{eqnarray*}
Hence, the generating functional can be expressed in terms of a series of free moments.

Now we apply this idea to the example nonlinear SDE
\begin{eqnarray*}
\dot{x}=-ax +bx^2 + y\delta(t-t_0) + \sqrt{D}x^{\frac{n}{2}}\eta(t) 
\end{eqnarray*}
for some $n\ge 0$.  For example, $n=0$ corresponds to standard additive noise (as in the OU process), while $n=1$ gives multiplicative noise with variance proportional to $x$.
The action for this equation is 
\begin{eqnarray}
S[x,\tilde{x}] &=& \int   dt \tilde{x}(\dot{x}+ax - bx^2 - y\delta(t-t_0))  - \tilde{x}^2 x^n\frac{D}{2}  \nonumber \\
&\equiv& S_F[x,\tilde{x}]  -y\tilde{x}(t_0) - b\int   dt \tilde{x}(t)x^2(t)   -\int dt \tilde{x}^2 x^n \frac{D}{2}
\label{action1}
\end{eqnarray}
where we have implicitly defined the ``free" action $S_F[x,\tilde{x}] = \int   dt \tilde{x}(\dot{x}+ax)$.     
Expectations with respect to this free action are
\begin{eqnarray*}
	\langle x(t) \tilde{x}(t') \rangle_F = G(t, t')
\end{eqnarray*}
where the propagator obeys
\begin{eqnarray*}
	 \left ( \frac{d}{dt} + a \right ) G(t,t') = \delta( t - t') 
\end{eqnarray*}
and all other moments are zero.
The generating functional is
\begin{eqnarray*}
Z[J,\tilde{J}]=\int {\cal D} x{\cal D}\tilde{x} e^{-S_F[x,\tilde{x}] + \int  \tilde{x} bx^2  \,dt +\int \tilde{x}y\delta(t-t_0) \,dt +\int \tilde{x}^2x^n \frac{D}{2} \, dt+\int \tilde{J}x dt + \int J \tilde{x} dt}
\end{eqnarray*}
The Taylor expansion of the exponential around the free action gives
\begin{eqnarray*}
Z[J,\tilde{J}]&=&\int {\cal D} x{\cal D}\tilde{x} e^{-S_F[x,\tilde{x}] }\left ( 1 + b\int  \tilde{x} x^2  \,dt + \tilde{x}(t_0)y+\frac{D}{2}\int \tilde{x}^2 x^n  \, dt+\int \tilde{J}x dt + \int J \tilde{x} dt \right. \\
&+& \left. 
\frac{1}{2!}\left (b\int  \tilde{x} x^2  \,dt + \tilde{x}(t_0)y +\frac{D}{2}\int \tilde{x}^2 x^n \, dt+\int \tilde{J}x dt + \int J \tilde{x} dt\right)^2 + \cdots\right)
\end{eqnarray*}

Because the free action $S_F$ is bilinear in $\tilde{x}, x$, the only surviving terms in the expansion are those with equal numbers of $x$ and $\tilde{x}$ factors.  Also, because of the Ito condition, $H(0)=0$, these pairings must come from \emph{different} terms in the expansion, e.g. the only term surviving from the first line is the very first term, regardless of the value of $n$.  All other terms come from the quadratic and higher terms in the expansion.  For simplicity in the remainder of this example we limit ourselves to $n=0$.  Hence, the expansion includes terms of the form
\begin{eqnarray*}
\lefteqn{Z[J,\tilde{J}]=\int {\cal D} x{\cal D}\tilde{x} e^{-S_F[x,\tilde{x}] }\left ( 1 \right.} \\
&+&
\frac{1}{2!}2\left (b\int \tilde{x}x^2 \tilde{x}(t_0)y\, dt  +b\int  \tilde{x} x^2  \,dt \int J \tilde{x} dt+\int \tilde{J}x  \tilde{x}(t_0) y\,dt + \int \tilde{J}x dt\int J \tilde{x} dt\right)\\
&+&
\frac{1}{3!}\frac{3!}{2!}b^2\frac{D}{2}\int \tilde{x} x^2 dt \int\tilde{x}x^2 dt\int\tilde{x}^2 dt\\
&+& 
\frac{1}{3!}\frac{3!}{2!}\frac{D}{2}\int \tilde{x}^2 \, dt  \int\tilde{J}x dt \int \tilde{J}x dt
+\frac{1}{3!}3!b\frac{D}{2}\int \tilde{x}x^2 \, dt  \int\tilde{x}^2 dt \int \tilde{J}x dt\\
&+&
\frac{1}{4!}\frac{4!}{2!}b\int \tilde{x}x^2 \, dt (\tilde{x}(t_0) y)^2 \int \tilde{J}x dt
+\frac{1}{4!}\frac{4!}{2!2!} (\tilde{x}(t_0) y)^2 \int \tilde{J}x dt\int \tilde{J}x dt\\
&+&
\left. 
\frac{1}{5!}5!b\frac{D}{2}\int \tilde{x}x^2 \, dt \int \tilde{x}^2 dt \tilde{x}(t_0) y \int \tilde{J}x dt\int\tilde{J}x dt+ \cdots\right)
\end{eqnarray*}
Note that this not an exhaustive list of terms up to fifth order.  Many of these terms will vanish because of $G(t,t') \propto H(t-t')$ and $H(0) =0$.
The combinatorial factors arise from the multiple ways of combining terms in the expansion.  There are $n!$ ways of combining terms at order $n$ and terms with $m$ repeats are divided by a factor of $m!$.
Completing the Gaussian integrals using Wick's theorem then yields
\begin{eqnarray}
\lefteqn{Z[J,\tilde{J}]= Z_F[0,0](1 }\nonumber\\
&+&
y\int G(t_1,t_0)\tilde{J}(t_1)\,dt_1 + \int \tilde{J}(t_1)G(t_1,t_2)J (t_2)dt_1dt_2 \nonumber\\
&+&
D\int G(t_2,t_1)G(t_3,t_1)\tilde{J}(t_2)\tilde{J}(t_3) \, dt_1 dt_2 dt_3\nonumber\\
&+& 
bD\int G(t_1,t_2)^2G(t_3,t_1) \tilde{J}(t_3) dt_1 dt_2 dt_3 \nonumber \\
&+&
by^2\int G(t_1,t_0)^2 G(t_2,t_1) \tilde{J}(t_2) dt_1 dt_2 
\nonumber\\
&+&
 y^2\int G(t_1,t_0)  \tilde{J}(t_1) \, dt_1\int G(t_2,t_0)  \tilde{J}(t_2) \, dt_2\nonumber\\
&+&
2bDy\int G(t_1,t_2)G(t_1,t_0)G(t_3,t_1)G(t_4,t_2)\tilde{J}(t_3)\tilde{J}(t_4)\,dt_1dt_2dt_3dt_4
+ \cdots )
\label{expansion}
\end{eqnarray}
As above, we have $Z_F[0,0]=1$. 

The moments and cumulants are obtained from (\ref{moment1}) and (\ref{cumulant1}) respectively.  For example, the mean is given by
\begin{eqnarray}
\lefteqn{\langle x(t) \rangle = \frac{1}{Z[0,0]}\frac{\delta}{\delta \tilde{J}(t)} Z[J(t),\tilde{J}(t)]_{J=0,\tilde{J}=0}} \nonumber \\
&=&yG(t,t_0) + bD\int G(t,t_1) G(t_1,t_2)^2  \, dt_1 dt_2 + b y^2 \int G(t,t_1) G(t_1,t_0)^2  \, dt_1 +\cdots
\label{mean}
\end{eqnarray}
The covariance is 
\begin{eqnarray*}
\lefteqn{\langle x(s)x(t) \rangle =  \frac{\delta}{\delta \tilde{J}(s)} \frac{\delta}{\delta \tilde{J}(t)} Z[J(t),\tilde{J}(t)]_{J=0,\tilde{J}=0}}\\
&=& D\int G(s,t_1)G(t,t_1) \, dt_1+ y^2 G(s,t_0)G(t,t_0)\\
 &+&
  2bDy  \int G(t_1,t_2)G(t_1,t_0)G(s,t_1)G(t,t_2) \, dt_1 dt_2 \nonumber \\
  &+&   2bDy  \int G(t_1,t_2)G(t_1,t_0)G(t,t_1)G(s,t_2) \, dt_1 dt_2
  \cdots
  \label{covariance}
\end{eqnarray*}
The first cumulant is the same as the mean but the second cumulant or covariance is
\begin{eqnarray}
\lefteqn{\langle x(s)x(t) \rangle_C =  \frac{\delta}{\delta \tilde{J}(s)} \frac{\delta}{\delta \tilde{J}(t)} \ln Z[J(t),\tilde{J}(t)]_{J=0,\tilde{J}=0}} \nonumber \\
&=&\frac{1}{Z}
 \left.\frac{\delta}{\delta \tilde{J}(s)} \frac{\delta}{\delta \tilde{J}(t)} Z\right|_{J=0,\tilde{J}=0}- \left.\frac{\delta}{\delta \tilde{J}(s)} Z\frac{\delta}{\delta \tilde{J}(t)} Z\right|_{J=0,\tilde{J}=0} \nonumber \\
&=& D\int G(s,t_1)G(t,t_1) \, dt_1\nonumber \\
 &+&
  2bDy  \int G(t_1,t_2)G(t_1,t_0)G(s,t_1)G(t,t_2) \, dt_1 dt_2 \nonumber \\
  &+&
  2bDy  \int G(t_1,t_2)G(t_1,t_0)G(t,t_1)G(s,t_2) \, dt_1 dt_2
  \cdots
  \label{eq:connectedcovariance}
\end{eqnarray}

 As can be seen in this example, the terms in the perturbation series become rapidly unwieldy.   However, a convenient means to keep track of the terms is to use Feynman diagrams,
which are graphs with edges connected by vertices that represents each term in the expansion of a moment.  The edges and vertices represent terms (i.e. interactions) in the action and hence SDE, which are combined according to a set of rules that reproduces the perturbation expansion shown above.  These are directed graphs (unlike the Feynman diagrams usually used for equilibrium statistical mechanics or particle physics). The flow of each graph, which represents the flow of time, is directed from right to left, points to the left being considered to be at times after points to the right.  The vertices represent points in time and separate into two groups:  \emph{endpoint} vertices and \emph{interior} vertices.  The moment $\left \langle \prod_{i=1}^N x(t_i) \prod_{j=1}^M \tilde{x}(t_j) \right \rangle$ is represented by diagrams with $N$ \emph{final} endpoint vertices which represent the times $t_i$ and $M$ \emph{initial} endpoint vertices which represent the times $t_j$.  Interior vertices are determined from terms in the action.

Consider the
interacting action expressed as the power series
\begin{eqnarray}
	S_I = \sum_{n\ge 2,m \ge 0}  V_{nm} = \sum_{n\ge 2,m \ge 0} \frac{v^{nm}}{n!}\int_{t_0}^\infty dt \tilde{x}^n x^m
	\label{eq:interactingAction}
\end{eqnarray}
where $n$ and $m$ cannot both be $\le 1$ (those terms are part of the free action).  (Nonpolynomial functions in the action are expanded in a Taylor series to obtain this form.)  
There is a vertex type associated with each $V_{nm}$.  The moment $\left \langle \prod_{i=1}^N x(t_i) \prod_{j=1}^M \tilde{x}(t_j) \right \rangle$ is given by a perturbative expansion of free action moments that are  proportional to $\left \langle \prod_{i=1}^N x(t_i) \prod_{j=1}^M \tilde{x}(t_j) V(N_v) \right \rangle_F$ where $V(N_v)$ represents a product of $N_v$ vertices.  Each term in this expansion corresponds to a graph with $N_v$ interior vertices. 
We label the $k$th vertex with time $t_k$.  As indicated in equation~(\ref{eq:interactingAction}), there is an integration over each such interior time point, over the interval $(t_0, \infty)$. The interaction $V_{nm}$ produces vertices with $n$ edges to the left of the vertex (towards increasing time) and $m$ edges to the right of the vertex (towards decreasing times).  Edges between vertices are due to Wick's theorem, which tells us that every $\tilde{x}(t')$ must be joined by a factor of $x(t)$ \emph{in the future}, i.e. $t>t'$, because $G(t,t') \propto H(t-t')$.  Also, by the Ito condition $H(0) = 0$, each edge must connect two \emph{different} vertices.  All edges must be connected, a vertex for the interaction $V_{nm}$ must connect to $n$ edges on the left and $m$ edges on the right.

\begin{figure}
	\includegraphics{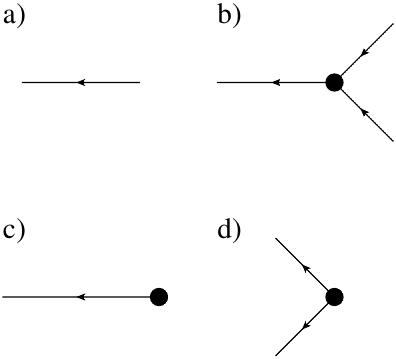}
	\caption{Feynman diagram components for a) an edge, the propagator $G(t,t')$, and vertices b) $\int b\tilde{x} x^2 dt$, c)  $\int y \tilde{x} \delta(t-t_0) dt$, and d) $\int\frac{D}{2}\tilde{x}^2 dt$.}
	\label{fig:feynman}
\end{figure}

\begin{figure}
	\includegraphics{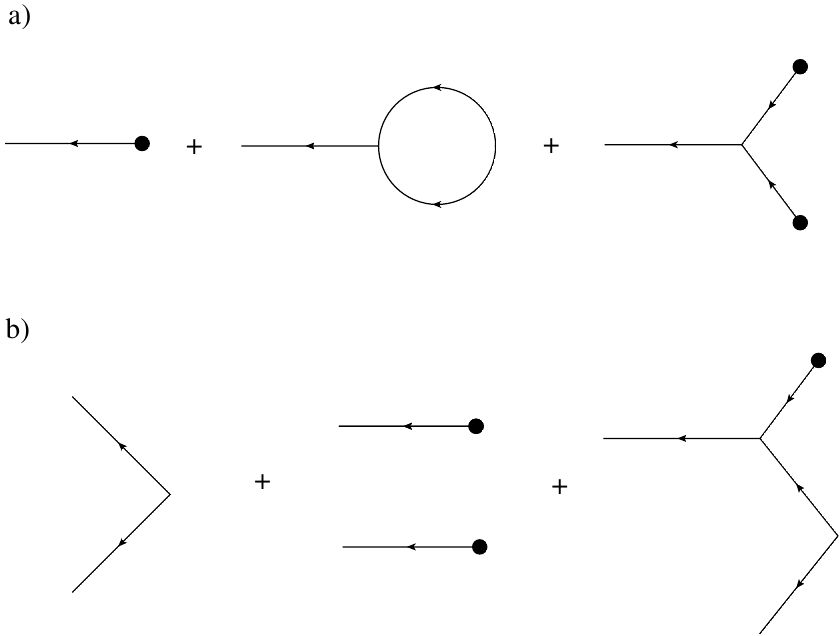}
	\caption{Feynman diagrams for a) the mean and b)  second moment.}
	\label{fig:mean2nd}
\end{figure}

Hence, terms at the $N_v$th order of the expansion for the moment $\left \langle \prod_{i=1}^N x(t_i) \prod_{j=1}^M \tilde{x}(t_j) \right \rangle$ are given by directed Feynman graphs with $N$ final endpoint vertices, $M$ initial endpoint vertices, and $N_v$ interior vertices with edges joining all vertices in all possible ways.  The sum of the terms associated with these graphs is the value of the moment to $N_v$th order.  Figure \ref{fig:feynman} shows the vertices applicable to action (\ref{action1}) with $n=0$.  Arrows indicate the flow of time, from right to left.  These components are combined into diagrams for the respective moments.  Figure~\ref{fig:mean2nd} shows three diagrams in the sum for the mean and second moment of $x(t)$.  The entire expansion for any given moment can be expressed by constructing the Feynman diagrams for each term.  

The application of Feynman diagrams for computing the diagrams corresponding to terms in a perturbative expansion is encapsulated in 
 the Feynman rules:

A)  For each vertex type, $V_{nm}$, which appears $k$ times in the diagram, there is a factor of $\frac{1}{k!}$.

B)  For $n$ distinct ways of connecting edges to vertices that yields the same diagram, i.e. the same topology, there is an overall factor of $n$.  This is the combinatoric factor from the number of different Wick contractions that yield the same diagram.

C)  Each vertex interaction $V_{nm}$ adds a factor $-\frac{v_{nm}}{n!}$.  The minus sign enters because the action appears in the path integral with a minus sign.

D)  For each edge between times $t$ and $t'$, there is a factor of $G(t,t')$.

E)  There is an integration over the times $t$ of each interior vertex over the domain $(t_0, \infty)$.

Comparing these rules with the diagrams in Figure~\ref{fig:mean2nd}, one can see the terms in the expansions in equations~(\ref{mean}) and (\ref{eq:connectedcovariance}), with the exception of the middle diagram in Figure~\ref{fig:mean2nd}b.  An examination of Figure~\ref{fig:mean2nd}a shows that this middle diagram is two copies of the first diagram of the mean.
Topologically, the diagrams have two forms.  There are connected graphs and disconnected graphs.  The disconnected graphs represent terms that can be completely factored into a product of moments of lower order (cf.~the middle diagram in Figure~\ref{fig:mean2nd}b).  Cumulants consist only of connected graphs since the products of lower ordered moments are subtracted from the moment by definition.  Thus, moments and cumulants can be computed directly from the diagrams that represent them.  
 The connected diagrams in Figure \ref{fig:mean2nd} lead to the expressions (\ref{mean}) and (\ref{eq:connectedcovariance}).
In the expansion (\ref{expansion}), the terms that do not include the source factors $J$ and $\tilde{J}$  only contribute to the normalization $Z[0,0]$ and do not affect moments because of (\ref{moment1}). Borrowing terminology from  quantum theory, these terms are called vacuum graphs and consist of closed graphs, i.e. they have no initial or trailing  edges.  In the cases we consider, all of these terms are $0$, which implies $Z[0,0] = 1$.

The diagrammatic expansion is particularly useful if the series can be truncated so that only a few diagrams need to be computed.  There are two types of expansions depending on whether the nonlinearity is small or the noise source is small.  In quantum theory, the small nonlinearity expansion is called a weak coupling expansion and the small fluctuation expansion is called a semiclassical or loop expansion.  The weak coupling expansion is straightforward.  Suppose one or more of the vertices is associated with a small parameter $\alpha$.  These vertices define the interacting action $S_I$ as demonstrated above.  Each appearance of that particular vertex diagram contributes a factor of $\alpha$ and the expansion can be continued to any order in $\alpha$.  

For the loop expansion, let us introduce the factor $h$ into the generating functional:
\begin{eqnarray}
	Z[J, \tilde{J}] = \int {\cal D} x(t){\cal D}\tilde{x}(t) e^{-\frac{1}{h}\left(S[x(t), \tilde{x}(t)] - \int \tilde{J}(t)x(t) dt - \int J(t) \tilde{x}(t) dt\right) }
	\label{eq:hgenfunc}
\end{eqnarray}
 According to the Feynman rules described above, with this change each diagram gains a factor of $h$ for each edge (internal or external) and a factor of $1/h$ for each vertex.  Let  $E$ be the number of external edges, $I$ the number of internal edges, and $V$ the number of vertices.  Then each connected graph now has a factor $h^{I+E-V}$.  It can be shown via induction that the number of closed loops $L$ in a given connected graph must satisfy $L = I - V + 1$ \cite{zinnjustin}.  To see this note that for diagrams without loops any two vertices must be connected by at most one internal edge.  Since the diagrams are connected we must have $V = I + 1$ when $L = 0$. Adding an internal edge between any two vertices increases the number of loops by precisely one.  Thus we see that the total factor for each diagram may be written $h^{E + L - 1}$.  We can organize the diagrammatic expansion in terms of the number of loops in the graphs.  This is called the loop expansion.  For the mean which has one external edge there are no factors of $h$ at lowest order.  Higher cumulants (which are determined by connected graphs) gain additional factors of $h$, e.g. the variance goes as $h$ at lowest order in the loop expansion.

 Loop diagrams arise because of nonlinearities in the SDE that couple to moments of the driving noise source.  For example, the middle graph in Figure~\ref{fig:mean2nd}a describes the coupling of the variance to the mean through the nonlinear $x^2$ term.  This produces a single loop diagram which is of order $h$, compared to the order $1$ ``tree" level mean graph.  Compare this factor of $h$ to that from the tree level diagram for the variance, which is order $h$.  This same construction holds for higher nonlinearities and higher moments for general theories.  The loop expansion is thus a series organized around the magnitude of the coupling of higher moments to lower moments.

As an example, consider the action
\begin{eqnarray*}
S[x,\tilde{x}]= \int  \tilde{x}(\dot{x}-f(x(t),t)) - \sigma^2 \frac{1}{2}\tilde{x}^2 g^2(x(t),t) \,dt
\end{eqnarray*}
where $\sigma$ is a small parameter and $f$ and $g$ are of order one.  Rescale the action with the transformation $\tilde{x} \rightarrow \tilde{x}/\sigma^2$ and $\tilde{J} \rightarrow  \tilde{J}/\sigma^2$.  The rescaled action now has the form
\begin{eqnarray*}
S[x,\tilde{x}]= \frac{1}{\sigma^2}\int  \tilde{x}(\dot{x}-f(x(t),t)) - \frac{1}{2}\tilde{x}^2 g^2(x(t),t) \,dt
\end{eqnarray*}
The generating functional is
\begin{eqnarray*}
Z[J,\tilde{J}]=\int {\cal D} x{\cal D}\tilde{x} e^{-\frac{1}{\sigma^2}\left(\int  \tilde{x}(\dot{x}-f(x(t),t)) - \frac{1}{2}\tilde{x}^2 g^2(x(t),t) \,dt +\int \tilde{J}x dt + \int J \tilde{x} dt \right ) }
\end{eqnarray*}
The loop expansion in this construction is explicitly a small noise expansion because $\sigma^2$ plays the role of $h$ in the loop expansion.
\begin{figure}
	\includegraphics{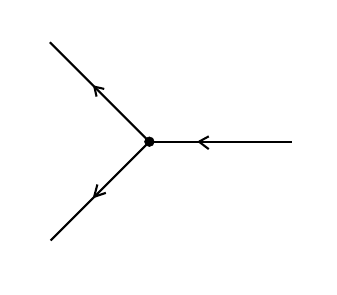}
	\caption{Vertex for multiplicative noise with $n=1$ in the action~(\ref{action1}).  This vertex replaces the one in Figure~\ref{fig:feynman}d.}
	\label{fig:feynmanMult}
\end{figure}

\begin{figure}
	\includegraphics{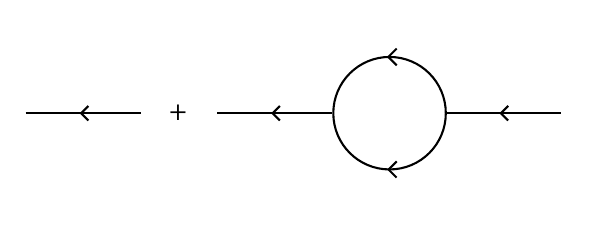}
	\caption{Feynam diagrams for the linear response, $\langle x(t) \tilde{x}(t') \rangle$, to one loop order.}
	\label{fig:loopProp}
\end{figure}

Now consider the one loop correction to the linear response, $\langle x(t) \tilde{x}(t') \rangle$, when $n=1$ in action~(\ref{action1}). For simplicity, we will assume the initial condition $y=0$. In this case, the vertex in Figure~\ref{fig:feynman}d now appears as in Figure~\ref{fig:feynmanMult}.  The linear response $\langle x(t) \tilde{x}(t') \rangle$ will be given by the sum of all diagrams with one entering edge and one exiting edge.  At tree level, there is only one such graph, equal to $G(t,t')$.  At one loop order, we can combine the vertices in Figures~\ref{fig:feynman}b and \ref{fig:feynman}d to get the second graph shown in Figure~\ref{fig:loopProp} to obtain
\begin{eqnarray*}
	\langle x(t) \tilde{x}(t') \rangle &=& G(t,t') + bD \int dt_1 dt_2 G(t,t_2) G(t_2, t_1)^2G(t_2, t') \\
	&=& e^{-a(t-t')}H(t - t') \left [ 1+ \frac{t - t'}{a} + \frac{1}{a^2} \left (e^{-a(t-t')} - 1 \right )  \right ]
\end{eqnarray*}

This loop correction arises because of two types of vertices.   There are vertices that we call ``branching" (as in Figure~\ref{fig:feynmanMult}), which have more exiting edges then entering edges.  The opposite case occurs for   those vertices which we call ``aggregating".   Noise terms in the SDE produce vertices with more than one exiting edge. As can be seen from the structure of the Feynman diagrams, all moments can be computed exactly when the deterministic part of the SDE is linear because it only involves convolving the propagator (i.e. Green's function) of the deterministic part of the SDE with the driving noise term, as in the case of the OU process above. On the other hand, nonlinearities give rise to vertices with more than one entering edge.  

\begin{figure}
	 \scalebox{0.7}{\includegraphics{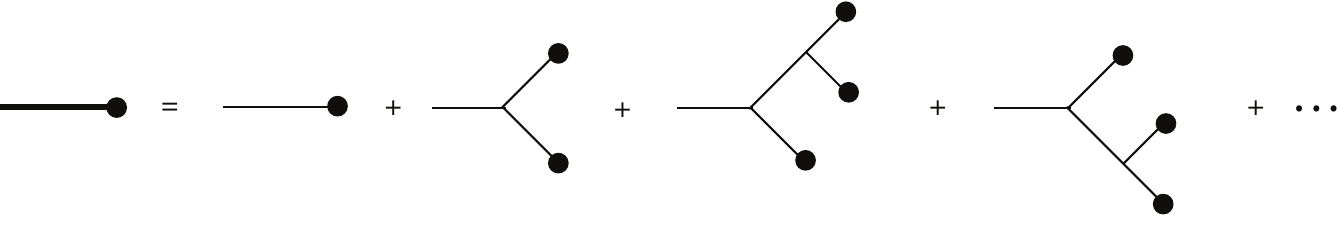}}
	\scalebox{0.7}{\includegraphics{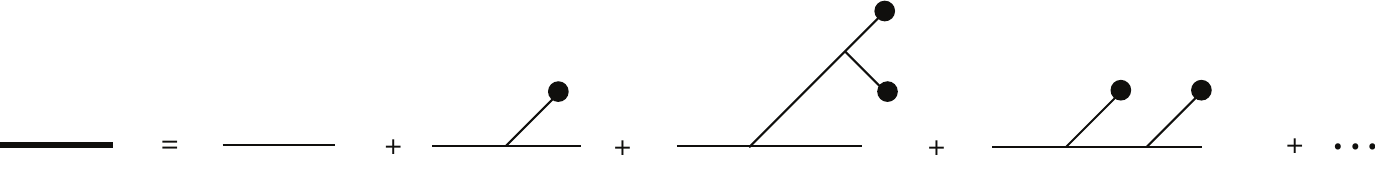}}
	\caption{Bold edges represent the sum of all tree level diagrams contributing to that moment.  Top) the mean $\langle x(t) \rangle_{\rm tree}$.  Bottom)  linear response $G_{\rm tree}(t,t')$.}
	\label{fig:classicalLines}
\end{figure}

If there are no branching vertices in the action (i.e. terms quadratic or higher in $\tilde{x}$), we do not even have an SDE at all, but just an ordinary differential equation. 
Consider the expansion of the mean for action~(\ref{action1}) for the case where $D=0$ (so that there is no noise term).  From equation~(\ref{mean}), we have
\begin{eqnarray*}
 	\langle x(t) \rangle  &=&yG(t,t_0) + b y^2 \int G(t,t_1) G(t_1,t_0)^2  \, dt_1 +\cdots
\end{eqnarray*}
The expansion for $D=0$ will be the sum of all tree level diagrams.  It is easy to see that in general this expansion will be the perturbative expansion for the solution of the ordinary differential equation obtained by discarding the stochastic driving term.  In other  words, the sum of the tree level diagrams for the mean satisfies
\begin{eqnarray}
	\frac{d}{dt} \langle x(t) \rangle_{\rm tree} = -a \langle x(t) \rangle_{\rm tree}+ b \langle x(t) \rangle_{\rm tree}^2
	\label{eq:meanfield}
\end{eqnarray}
along with the initial condition $\langle x(t_0) \rangle_{\rm tree} = y$.   Similarly, the sum of the tree level diagrams for the linear response, $\langle x(t) \tilde{x}(t') \rangle_{\rm tree} = G_{\rm tree}(t,t')$, is the solution of the linearization of (\ref{eq:meanfield}) with a Dirac delta functional initial condition, i.e. the propagator
\begin{eqnarray*}
	\frac{d}{dt} G_{\rm tree}(t,t')= -a G_{\rm tree}(t,t')+ 2b \langle x(t) \rangle_{\rm tree}G_{\rm tree}(t,t') + \delta(t - t')
\end{eqnarray*}
 The semiclassical approximation amounts to a small noise perturbation around the solution to this equation.
We can represent the sum of the tree level diagrams graphically by using bold edges, which we call  ``classical" edges, as in Figure~\ref{fig:classicalLines}.   We can then use the classical edges within the loop expansion to compute semiclassical approximations to the moments of the solution to the SDE.  The one loop semiclassical approximation of the mean for the case $n=0$ is given by the sum of the first two graphs in Figure~\ref{fig:mean2nd}a with the thin edges replaced by bold edges.  For the covariance, the first graph in Figure~\ref{fig:mean2nd}b suffices, again with thin edges replaced by bold edges.  
These graphs are equivalent to the equations:
\begin{eqnarray}
	\langle x(t) \rangle &=& \langle x(t) \rangle_{\rm tree} + b D \int_{t_0}^{t} dt_1 \int_{t_0}^{t_1} dt_2 G_{\rm tree}(t,t_2) G_{\rm tree}(t_2,t_1)^2  
	\label{meantree}
\end{eqnarray}
and
\begin{eqnarray}
	\langle x(t) x(t') \rangle &=& D \int_{t_0}^{{\rm min}(t,t')} dt_1 G_{\rm tree}(t, t_1) G_{\rm tree}(t',t_1)
	\label{2ndtree}
\end{eqnarray}
Using equation (\ref{2ndtree}) in  (\ref{meantree}) gives 
\begin{eqnarray*}
	\langle x(t) \rangle &=& \langle x(t) \rangle_{\rm tree} + b D \int_{t_0}^{t} dt_1  G_{\rm tree}(t,t_2)   \langle x(t_2) x(t_2) \rangle
\end{eqnarray*}
 This approximation is first order in the dummy loop parameter $h$ for the mean (one loop) and covariance (tree level).  
For the case $n=1$, equations (\ref{meantree}) and (\ref{2ndtree}) are
\begin{eqnarray*}
	\langle x(t) \rangle &=& \langle x(t) \rangle_{\rm tree} + b D \int_{t_0}^{t} dt_1 \int_{t_0}^{t_1} dt_2 G_{\rm tree}(t,t_2) G_{\rm tree}(t_2,t_1)^2  \langle x(t) \rangle_{\rm tree}
\end{eqnarray*}
and
\begin{eqnarray*}
	\langle x(t) x(t') \rangle &=& D \int_{t_0}^{{\rm min}(t,t')} G_{\rm tree}(t, t_1) G_{\rm tree}(t',t_1)\langle x(t) \rangle_{\rm tree}
\end{eqnarray*}
 Using the definition of $G_{\rm tree}(t,t')$,  the self-consistent semiclassical approximation  for $\langle x(t) \rangle$ to one-loop order is
 \begin{eqnarray*}
 	 \frac{d}{dt}\langle x(t) \rangle  + a\langle x(t) \rangle  - b \langle x(t) \rangle^2 &=&  b D \int_{t_0}^{t} dt_1 G_{\rm tree}(t,t_1)^2  \langle x(t) \rangle
 \end{eqnarray*}
 or
  \begin{eqnarray*}
 	 \frac{d}{dt}\langle x(t) \rangle  + a\langle x(t) \rangle  - b \langle x(t) \rangle^2 &=&  b  \int_{t_0}^{t} dt_1 \langle x(t_1) x(t_1) \rangle_C
 \end{eqnarray*}
The semiclassical approximation known as the ``linear noise" approximation takes the tree level computation for the mean and covariance.  The formal way of deriving these self-consistent equations is via the \emph{effective action}, which is beyond the scope of this review.  We refer the interested reader to \cite{zinnjustin}.


\section{Connection to Fokker-Planck equation}
\label{sec:fp}

In stochastic systems, one is often interested in the PDF $p(x,t)$, which gives the probability density of position $x$ at time $t$.  This is in contrast with the probability density functional $P[x(t)]$ which is the probability density of all possible functions or paths $x(t)$. Previous sections have been devoted to computing the moments of $P[x(t)]$, which provide the moments of $p(x,t)$ as well.  In this section we leverage knowledge of the moments of $p(x,t)$ to determine an equation it must satisfy.  In simple cases, this equation is a Fokker-Planck equation for  $p(x,t)$.  

The PDF $p(x,t)$ can be formally obtained from $P[x(t)]$ by marginalizing over the interior points of the function $x(t)$.  Consider the
transition probability $U(x_1,t_1 | x_0, t_0)$ between two points $x_0,t_0$ and $x_1,t_1$.  This is equal to $p(x,t)$ given the initial condition $p(x,t_0) = \delta(x - x_0)$.  In terms of path integrals this can be expressed as
\begin{eqnarray*}
U(x_1,t_1|x_0,t_0) = \int^{(x(t_1)=x_1)}_{(x(t_0)=x_0)} {\cal D}x(t) \, P[x(t) ]
\end{eqnarray*}
where the upper limit in the integral is fixed at $x(t_1)=x_1$ and the lower at $x(t_0)=x_0$. The lower limit appears as the initial condition term in the action and can thus be considered part of $P[x(t)]$. The upper limit on the path integral can be imposed with a functional Dirac delta via 
\begin{eqnarray*}
U(x_1,t_1|x_0,t_0) = \int {\cal D}x(t) \, \delta (x(t_1)-x_1)P[x(t)]
\end{eqnarray*}
which in the Fourier representation is given by
\begin{eqnarray*}
U(x_1,t_1|x_0,t_0) = \frac{1}{2\pi i}\int d\lambda\int {\cal D}x(t) \,  e^{\lambda (x(t_1)-x_1)}P[x(t)]
\end{eqnarray*}
where the contour for the $\lambda$ integral runs along the imaginary axis. This can be rewritten as
\begin{eqnarray}
U(x_1,t_1|x_0,t_0) = \frac{1}{2\pi i}\int d\lambda \, e^{-\lambda(x_1-x_0)} Z_{\rm CM}(\lambda)
\label{transition}
\end{eqnarray}
in terms of an initial condition centered moment generating function
\begin{eqnarray*}
Z_{\rm CM}(\lambda) = \int {\cal D}x \,  e^{\lambda (x(t_1)-x_0)}P[x(t)]
\end{eqnarray*}
where the measure ${\cal D}x(t)$ is defined such that $Z_{\rm CM}(0)=1$.  Note that this generating function $Z_{\rm CM}(\lambda)$ is different from the generating functionals we presented in previous sections.  $Z_{\rm CM}(\lambda)$ generates moments of the deviations of $x(t)$ from the initial value $x_0$ at a specific point in time $t$.
Taylor expanding the exponential gives
\begin{eqnarray*}
Z_{\rm CM}(\lambda) = 1+\sum_{n=1}^\infty \frac{1}{n!}\lambda^n \langle (x(t_1)-x_0)^n\rangle_{x(t_0)=x_0}
\end{eqnarray*}
where
\begin{eqnarray*}
\langle (x(t_1)-x_0)^n\rangle_{x(t_0)=x_0} = \int {\cal D}x\, (x(t_1)-x_0)^nP[x(t)] 
\end{eqnarray*}
Inserting into (\ref{transition}) gives
\begin{eqnarray*}
U(x_1,t_1|x_0,t_0) = \frac{1}{2\pi i }\int d\lambda \, e^{-\lambda(x_1-x_0)} \left(1+\sum_{n=1}^\infty \frac{1}{n!}\lambda^n \langle (x(t_1)-x_0)^n\rangle\right)
\end{eqnarray*}
Using the identity
\begin{eqnarray*}
\frac{1}{2\pi i}\int d\lambda \, e^{-\lambda(x_1-x_0)} \lambda^n  = \left(-\frac{\partial}{\partial x_1} \right)^n\delta(x_1-x_0)
\end{eqnarray*}
results in
\begin{eqnarray}
U(x_1,t_1|x_0,t_0) = \left(1+\sum_{n=1}^\infty \frac{1}{n!}\left(-\frac{\partial}{\partial x_1}\right)^n \langle (x(t_1)-x_0)^n\rangle_{x(t_0)=x_0}\right)\delta(x_1-x_0)
\label{trans2}
\end{eqnarray}

The probability density function $p(y,t)$ obeys 
\begin{eqnarray}
p(y,t+\Delta t) = \int U(x,t+\Delta t|y',t) p(y',t) dy'
\end{eqnarray}
Inserting (\ref{trans2}) gives
\begin{eqnarray*}
p(y,t+\Delta t) = \left(1+\sum_{n=1}^\infty \frac{1}{n!}\left(-\frac{\partial}{\partial y}\right)^n \langle (x(t+\Delta t)-y)^n\rangle_{x(t)=y}\right) p(y,t)
\end{eqnarray*}
 Expanding $p(y,t+\Delta t)$ and the moments  in a Taylor series in $\Delta t$ gives
 \begin{eqnarray*}
\frac{\partial p(y,t) }{\partial t} \Delta t= \sum_{n=1}^\infty \left(-\frac{\partial}{\partial y}\right)^n \frac{1}{n!}\langle (x(t+\Delta t)-y)^n \rangle_{x(t)=y} p(y,t) +O(\Delta t^2)
\end{eqnarray*}
since $x(t)=y$.  
In the limit $\Delta t\rightarrow 0$ we obtain the Kramers-Moyal expansion
 \begin{eqnarray*}
\frac{\partial p(y,t) }{\partial t} = \sum_{n=1}^\infty\frac{1}{n!} \left(-\frac{\partial}{\partial y}\right)^n D_n(y,t) p(y,t) +O(\Delta t^2)
\end{eqnarray*}
where the \emph{jump moments} are defined by
\begin{eqnarray}
D_n(y,t)=\lim_{\Delta t \rightarrow0}\left. \frac{\langle \left ( x(t+\Delta t) - y\right)^n\rangle}{\Delta t}
\right|_{x(t)=y}
\label{Dn}
\end{eqnarray}
As long as these limits are convergent, then it is relatively easy to see that only connected Feynman graphs will contribute to the jump moments.  In addition, we can define $z=x-y$,   $\tilde{z} = \tilde{x}$ and use the action $S[z(t)+y, \tilde{z}(t)]$.  This shift in $x$ removes the initial condition term.  This means we can calculate the $n$th jump moment by using this shifted action to compute the sum of all graphs with no initial edges and $n$ final edges (as in Figure~\ref{fig:feynman}d for $n=2$).

As an example, consider the Ito SDE (\ref{sde}).  From  the discretization (\ref{discrete}), where $h=\Delta t$, it is found that
\begin{eqnarray}
\lim_{\Delta t \rightarrow0}\left. \frac{\langle \left ( x(t+\Delta t) - y \right )^n\rangle}{\Delta t}\right|_{x(t)=y} = \lim_{\Delta t \rightarrow0}\frac{\left \langle \left ( f_i(y)\Delta t - g_i(y)w_i\sqrt{\Delta t}\right )^n\right \rangle}{\Delta t}
\end{eqnarray}
Which yields $D_1(y,t)= f(y,t)$, $D_2=g(y,t)^2$ and $D_n=0$ for $n>2$.  Thus for the Ito SDE (\ref{sde}), the Kramers-Moyal expansion becomes the
 the Fokker-Planck equation
\begin{eqnarray*}
\frac{\partial p(y,t) }{\partial t} =\left(-\frac{\partial}{\partial y} D_1(y,t) + \frac{1}{2}\frac{\partial^2}{\partial^2 y} D_2(y,t)\right)p(y,t)
\end{eqnarray*}
We have $D_n=0$ for $n>2$ even though there are non-zero contributions from connected graphs to these moments for $n>2$ in general. However, all of these moments require the repeated use of  the vertex with two exiting edges; this will cause $D_n \propto \Delta t^m$ for some $m > 1$ and thus the jump moment will be zero in the limit.

We can envision actions for more general  stochastic processes by considering vertices which have more than two exiting edges, i.e. we can add a term to the action of the form
\begin{eqnarray*}
	S_V[x,\tilde{x}] = \frac{1}{n!}\int dt \tilde{x}^n h(x) 
\end{eqnarray*}
for some $n$ and function $h(x)$.  This will produce a non-zero $D_n$.  The PDF for this kind of process will not in general be describable by a Fokker-Planck equation, but will need the full Kramers-Moyal expansion.  If we wished to provide an initial distribution for $x(t_0)$ instead of specifying a single point, we could likewise add similar terms to the action.  In fact, the completely general initial condition term is given by
\begin{eqnarray*}
	S_{\rm initial}[\tilde{x}(t_0)] = - \ln Z_{y}[\tilde{x}(t_0)] 
\end{eqnarray*}
where $Z_{y}$ is the generating functional for the initial distribution.  In other words, the initial state terms in the action are the cumulants of the initial distribution multiplied by the corresponding powers of $\tilde{x}(t_0)$.  

Returning to the Ito process (\ref{sde}), the solution to the Fokker-Planck equation can be obtained directly from the path integral formula for the transition probability (\ref{transition}).  
Let $\ln Z[\lambda]$ be the cumulant generating function for the moments of $x(t)$ at time $t$.  It can be expanded as
\begin{eqnarray*}
Z_{\rm CM}[\lambda]=\exp\left[\sum_{n=1}\frac{1}{n!}\lambda^n\langle x(t)^n \rangle_C\right]
\end{eqnarray*}
yielding
\begin{eqnarray*}
p(x,t) = \frac{1}{2\pi i}\int d\lambda \, e^{-\lambda x} \exp\left[\sum_{n=1}\frac{1}{n!}\lambda^n\langle x(t)^n \rangle_C\right]
\end{eqnarray*}




For the Ornstein-Uhlenbeck process the first two cumulents are given in (\ref{meanOU}) and (\ref{varOU}) yielding (assuming initial condition $x(t_0) = y$)

\begin{eqnarray}
p(x,t)=  \sqrt{\frac{a}{\pi D(1-e^{-2a(t-t_0)})}}\exp\left(\frac{-a(x-ye^{-a(t-t_0)})^2}{D(1-e^{-2a(t-t_0)}}\right)
\end{eqnarray}

\section{Further reading}

The reader interested in this approach is encouraged to explore the extensive literature on path integrals and field theory.  The reader should be aware that most of the references listed will concentrate on applications and formulations appropriate for equilibrium statistical mechanics and particle physics, which means that they will not explicitly discuss the response function approach we have demonstrated here.    For  application driven examinations of path integration there is Schulman\cite{schulman} and Kleinert\cite{2004piqm.book.....K}.  More mathematically rigorous treatments can be found in Simon\cite{simon} and Glimm and Jaffe \cite{glimm}.  For the reader seeking more familiarity with concepts of stochastic calculus such as Ito or Stratonovich integration there are applied approaches \cite{Gardiner:2004eu} and rigorous treatments \cite{karatzas} as well.  Zinn-Justin \cite{zinnjustin} covers a wide array of topics of interest in quantum field theory from statistical mechanics to particle physics.  Despite the exceptionally terse and dense presentation, the elementary material in this volume is recommended to those new to the concept of path integrals.  Note that Zinn-Justin covers SDEs in a somewhat different manner than that presented here (the Onsager-Machlup integral is derived; although see chapters 16 and 17), as does Kleinert.  We should also point out the parallel between the form of the action for exponential decay (i.e. $D = 0$ in the OU process) and the holomorphic representation of the harmonic oscillator presented in \cite{zinnjustin}.  The response function formalism was introduced by Martin, Siggia, Rose \cite{martin}.  Closely related path integral formalisms have been introduced via the work of Doi \cite{doi1,doi2} and Peliti \cite{peliti} which have been used in the analysis of reaction-diffusion system \cite{cardyrev,cardyrev2, Janssen2005147,tauber}.  Uses of path integrals in neuroscience have appeared in \cite{bc, buice-cowan,bcc,bressloffpath}.

\bibliography{pathreview.bib}
\end{document}